\documentclass[12pt,preprint]{aastex}

\shorttitle{QSO + galaxy association and discrepant redshifts in NEQ3}
\shortauthors{C. M. Guti\'errez and L\'opez-Corredoira}

\begin{document}

\title{QSO+Galaxy association and discrepant redshifts in NEQ3}

\author{C. M. Guti\'errez}
\affil{Instituto de Astrof\'{\i}sica de Canarias, E-38205, La Laguna, 
Tenerife, Spain}
\email{cgc@ll.iac.es}
\and
\author{M. L\'opez-Corredoira}
\affil{Astronomisches Institut der Universit\"at Basel.
Venusstrasse 7. CH-4102 Binningen, Switzerland}


\begin{abstract}

Spectroscopy and deep imaging of the group NEQ3 are presented. This
system is formed by three compact objects with relative separations of
$\sim$2.6 and $\sim$2.8 arcsec, and a lenticular galaxy at $\sim$17
arcsec from the geometric centre of the group. A diffuse filament is
located on a line joining the three compact objects and the main
galaxy. Analysis of these observations confirms the redshift previously
known for three of the objects ($z=0.1239$ for the main galaxy, and
$z=0.1935$ and 0.1939 for two of the compact objects). We have also
determined the previously unknown redshift of the third compact object
as $z=0.2229$. Using the relative strength and width of the main
spectral lines we have classified the compact objects as two HII
galaxies and one QSO (the object at $z=0.1935$). With cross-correlation
techniques, we have tentatively estimated the redshift of the filament
as $z=0.19$ (although a weaker component also appears at $z=0.12$) so
that it is probably associated with the halo of the two compact objects
at this redshift. The two objects at redshift $\sim 0.19$ represent
possibly one of the  more clear examples of starburst (and perhaps QSO
activity) driven by interaction.  These, and the relation between these
two objects and the other two at ($\sim 0.12$ and $\sim 0.22$) make the
nature of this system intriguing, being difficult to explain the whole
association on conventional  scenarios.

\end{abstract}


\section{Introduction}

In the last three years we have started a programme to study  several
apparently interacting systems with different redshifts. These
constitute the so-called problem of anomalous redshifts (e.g.\ Burbidge
1996 and references therein). The first system that we studied in
detail is formed by NGC 7603, its companion galaxy NGC 7603B, the
filament apparently connecting both galaxies, and two objects within it
(L\'opez-Corredoira \& Guti\'errez 2002). In that article we showed
that the two objects in the filament have redshifts of 0.24 and 0.39
respectively. In L\'opez-Corredoira \& Guti\'errez (2003) we have
presented subsequent observations on these objects and detected others
at high redshift in the same field. Simple probability computations
indicate the unlikelihood that this configuration is merely a
projection effect.

We continued our  study of several other systems in different observing
campaigns conducted during 2002. Here, we present our observations and
analysis of the system NEQ3 (the  designation was proposed by Sulentic
\& Arp 1976 and means Non Equilibrium System Number 3). Although a
rather intriguing system, the only other study of it, surprisingly, has
been conducted by Arp (1977). The system is centred at RA(J2000) =
12$^{\rm h}$ 09$^{\rm m}$ 51.6$^{\rm s}$, Dec.(J2000) = +39$^{\circ}$
31$^{\prime}$ 29$^{\prime\prime}$ at 10.8  arcmin from the exhaustively
studied large active galaxy NGC 4151. Other interesting associations
were found by Arp  in the region around this galaxy. The system
comprises three compact objects with relative separations of 2.6 and
2.8 arcsec with respect to the central object, all of them lying  along
the minor axis of an apparent lenticular galaxy  at $\sim$17 arcsec. 
Arp measured a redshift of 0.12 for the main galaxy  and 0.19 for two
of the compact objects. Arp described the objects as compact with
emission lines. A filament is situated along the line connecting the
main galaxy and the three compact objects.

All the above facts attracted our attention and made this system one of
our top priorities. The goals of our study were to ascertain: i) the
redshift of the third compact object, ii) the nature of these objects,
iii) the redshift of the mentioned filament, iv) whether there are
other objects with anomalous redshifts in the near field, v) whether
there is evidence for distortions in the profile of the lenticular
galaxy, vi) how likely it is that this optical association is merely a 
projection effect, and vii) what physical mechanisms could originate a
system like this.

In this article we present the first results of this study. A more
detailed analysis covering the photometry and velocity radial profile
of the main galaxy and spectra of other objects in the field is in
progress and will be presented in a future paper. 

\section{Observations}

The observations presented here comprise broad band imaging and
long-slit spectroscopy with intermediate resolution. These observations
were taken at the NOT\footnote {The Nordic Optical Telescope (NOT) is
jointly  operated on the island of La Palma by Denmark, Finland,
Iceland, Norway, and Sweden, in Spain's Roque de los Muchachos
Observatory of the Instituto de Astrof\'\i sica de Canarias (IAC).} (on
2002 December 2) and WHT\footnote {The William Herschel Telescope (WHT)
is operated by the Isaac Newton Group and the IAC in Spain's Roque de
los Muchachos Observatory.} (on 2002 December 28--29). We follow a
standard reduction method explained in L\'opez-Corredoira \&
Guti\'errez (2003). Table~1 presents a summary of the observations. We
observed the field in the Sloan  $r^\prime$ filter. We used the NOT
with the ALFOSC instrument, which has a field of view  $\sim$6 arcmin
with a spatial sampling of $0.188$ arcsecs. The $r^\prime$ band was
relatively calibrated using an image taken with the IAC 80 telescope
\footnote {The IAC-80 telescope is located at Spain's Teide Observatory
(Tenerife) and is operated by the IAC.} and corrected to the SDSS
$r^{\prime}$ filter through the relations given in Smith et al.\
(2002).

The spectra were taken with the red arm of the ISIS instrument and the
R158R grism at the WHT. This configuration gives a sampling of 1.63
\AA\ and a resolution,  measured from the width of the spectral arc
lines, of $\sim$8 \AA~for a slit width of 1.2 arcsec. The spectra have
a useful range between $\sim$3900 and $\sim$9600 \AA.

Figure~1 presents a false-colour scale of our image in the $r^\prime$
band. Superimposed are the positions in which we positioned the slits.
These positions were chosen to cross the three compact objects, the
main galaxy, the filaments within them and several other promising
objects in the field. For clarity, the objects analysed here have been
arbitrary labelled from 1 to 4.

\section{Analysis}

Figure~2 shows the system in detail. Objects 1, 2 and 3 appear rather 
compact, although two small distortions (see also Figure~1) in object
2, seem to point in the direction of the two close companions.
Unfortunately our ability to detect small-scale structure is severely
restricted by poor seeing; for example, the seeing (1.4 arcsec) in the
$r^\prime$ band corresponds to $\sim$4 kpc at a cosmological redshift
of $z=0.19$. We see how the system is surrounded by a diffuse emission.
We have delineated this emission down to isophote 26.8 mag
arcsec$^{-2}$. We also discern a filament (previously noted by Arp)
along the line of the minor axis of object 4. It seems also that the
filament decreases in brightness from object 1--2 to object 4, which
looks rather symmetric with a morphological shape typical of a
lenticular (S0/Sa) galaxy, its spectrum presenting only  absorption
lines characteristic of an old population (Ca H, Ca K, MgI
($\lambda 5180$), NaI($\lambda 5892$)). Using them we have estimated a
redshift of  $0.1239\pm 0.0005$ for this galaxy.  The only signs of
distortion are the filament in the direction to objects 1--3, and an
extended emission feature to the north that seems to enclose a weak
object ($r^\prime=21.8$ mag) situated at $\sim 13$ arcsec.  We have 
determined the magnitudes of the four objects using the Sextractor
software (Bertin \& Arnouts 1996). The magnitudes of objects 1--4 in
the $r^\prime$ band are respectively 19.8, 19.6, 20.2 and 17.3.  As a
consequence of the proximity of the three objects and the existence of
diffuse emission around them, the uncertainty in the quoted magnitudes
should be $\sim$0.1--0.2 mag.

The spectra of objects 1, 2 and 3 are dominated by emission lines whose
identification allow us to determine their redshift unambiguously. The 
main features are the lines of the hydrogen Balmer series, and the
lines of OII and OIII. Other minor features such as  SI,  NI, NII and
HeI are also detected.  Figure~3, shows the spectra (corrected of
redshift)  around the lines H$\beta$ and  H$\alpha$. Table~2 presents
the equivalent widths of each of these lines. The positions of all of
these lines is consistent within the uncertainty in our spectral
resolution. The resulting redshifts are $0.1935\pm 0.0002$, $0.1939\pm
0.0005$ and $0.2229\pm 0.0002$ for  objects 1, 2 and 3 respectively.
Tentatively we have identified some possible absoption features in the
spectra of the three objects which could be associated to the host or
normal underlying stellar  population.

Object 1 has a typical broad band line spectrum, while objects 2 and 3
have only narrow emission lines. They have strong star formation
(EW(H$\alpha$ + NII) = 100 and 70 m\AA~respectively; Carter et al.\
2001). We have classified the emission line objects using the
diagnostics considered by Ver\'on, Gon\c calves, \& Ver\'on-Cetty
(1997) based on the line ratios of OIII($\lambda$5007)/H$\beta$ and
NII($\lambda$6584)/H$\alpha$.  Because of the spectral proximity of the
lines involved in each ratio, the effects of reddening are minimized
and the unaccuracies on the determination of the continuum largely
cancel out. According to these criteria, object 1 is a QSO/Seyfert 1
galaxy, while objects 2 and 3 are HII galaxies.

The spectrum of the filament is very noisy and does not have any
obvious feature. However, using the subroutine $fxcor$ of
IRAF\footnote{IRAF is the Image Reduction and Analysis Facility,
written and supported by the IRAF programming group at the National
Optical Astronomy Observatories (NOAO) in Tucson, Arizona.} with a
field bright star as template, we tentatively identify a maximum in the
cross-correlation function which corresponds approximately to a
redshift of 0.19. A secondary maxima appears also at $z=0.12$, but we
do not exclude possibly contaminnation from the main galaxy.

\section{Discusion}

As in other cases of anomalous redshift, the optical configuration of
the system formed by NEQ3 (objects 1 to 3), the close lenticular galaxy
(object 4) and the additional features such as the filament apparently
connecting them, and the diffuse emission in the northern direction
seem to be clear indications of proximity and interaction. As discussed
in L\'opez-Corredoira \& Guti\'errez (2003), examples like this, in
which galaxies interact through filaments and show distortions in the
halos, are relative common. An interpretation that explains the
configuration as equivalent to other systems in interaction would be
clearly preferred over one in which the configuration is purely a
projection effect.

Objects 1 and 2 have redshifts of $z=0.1935$ and 0.1939 respectively
and are separated by 2.8 arcsec.  Considering the redshift as
indicative of cosmological distance, the above numbers are translated
into a difference of $\sim$100 $\pm$ 200 km s$^{-1}$ in velocity and a
separation of $\sim$8 kpc. Bahcall et al.\ (1997), in a study with the
{\itshape Hubble Space Telescope}, found that most of the detected QSO
hosts have a galaxy at less than 25 kpc. This companion tends to be an
elliptical galaxy for radio-loud QSOs, and an elliptical or spiral
galaxy for radio-quiet QSOs. Canalizo \& Stockton (2001) have studied
the environment of nine objects  with intermediate colors between ultra
luminous infrared galaxies and QSOs and have found close companion in
some of them and morphological evidence of  interaction in eight of
them. In general the companion show a mixture of young and old stellar
population but are not star forming galaxies. In that sample one of the
system which perhaps resemble more NEQ3 is  IRAS00275-2859 which  shows
a secondary nucleus (or a giant HII region)  at $\sim 2.7$ arcsec from
the QSO and a tail with several clumps extending 12 arcsec. Differently
from the cases studied by these authors, in NEQ3 the two objects (the
QSO and the HII galaxy) at redshift $\sim 0.19$ have similar magnitudes
in the $r$ band. 

The association of objects 1 and 2 is particularly interesting because
it is one of the cases with a smaller angular separation, the galaxy is
undergoing an intense burst of star formation, and there is a diffuse
filament possibly associated with this pair. All of these facts seem to
indicate a strong interaction between the QSO and the HII galaxy.  This
interaction could be the responsible of the intense star formation in
object 2 and perhaps also of the QSO activity (Stockton 1982).

In a conventional scenario the role of objects 3 and 4 is unclear. For
instance, could the difference in redshift between objects 1 and 2, and
object 3  ($\sim$0.03) be produced by a difference in peculiar
velocity? The difference would be 9000 km s$^{-1}$. As far as we know,
interactions between galaxies with such differences in velocity have
not been observed and it would be difficult to explain in the framework
of models of galaxy formation. In favour of the physical relation
between the pair of objects 1--2  and object 3 is the main halo which
seems surround the three objects uniformily and the slight elongation 
(see Figures 1 and 2) of object 2 through objects 1 and 3. 

We have discarded the possibility of gravitational lensing produced by
object 4, because the combination of amplification and separation
between the sources and the lens would require much more massive
lensing structures such as the cores of galaxy clusters. This
amplification by galaxy clusters has been used by Ellis et al. 
(2001) to detect high redshift galaxies. 

Finally, is it just a merely projection effect the proximity of object
4? Is this proximity related with the presence of the possible filament
extending outwards in the direction of the three compact objects? From
the galaxy counts by Metcalfe et al. \ (1991) and assuming Poission
statistics, we have made a simple probability calculation. Based on
that, Table~3 presents an estimation of the integrated number of 
galaxies  with magnitudes up the one of objects 2, 3 and 4. We start
the computation  with the pair of objects at $z\sim 0.19$, and  compute
the probability ($P_1$) to find an object  as bright as 3 (which
density is $N_3$ in a circle of $\sim 3$ arcsec of radius (area $A_1$).
Assuming Poisson statistics (and onsidering that the density of each
kind of objects in the area considered is  very low) we have
$P_1=N_3\times A_1\sim 4\times 10^{-3}$.  The probability to have
object 4 at the end of the filament (assuming roughly an area for this
filament $A_2\sim 15\times 2.5=37.5$  arcsec$^{-2}$)  is $P_2=N_4\times
A_2\sim 5\times 10^{-4}$. Then the total probability is:
$P=P_1*P_2/2\sim 10^{-6}$. This low probability makes unlikely an
explanation based in merely a projection effect of the four objects.

\section{Summary}

\begin{itemize}

\item New observations of a system of anomalous redshift are presented.
These comprises, broad band, and intermediate resolution  spectroscopy
of four close objects.

\item  The redshifts of the main galaxy ($0.1239\pm 0.0005$) and the
two previously known compact objects (0.1935 $\pm$ 0.0002 and 0.1939
$\pm$ 0.0002) have been confirmed with better accuracy.

\item We have delineated the diffuse emission apparently connecting the
main galaxy and the three compact objects up to the isophote 26.8 mag
arcsec$^{-2}$ in the SDSS  $r^{\prime}$ filter.

\item The redshift of the third compact object has been estimated as
$0.2229\pm 0.0002$.

\item We have shown that the filament probably has a redshift of
$z=0.19$ and seems to be mainly associated with the compact objects.

\item From the ratio of several emission lines, we have classified the
three emission line objects as a QSO (object 1) and  HII galaxies
(objects 2 and 3) 

\item The probability of the whole aggregation being fortuitous has been computed as  $\sim$$10^{-6}$.

\end{itemize}

\acknowledgments We want to thank Fernanda Art\'\i guez Carro, who
kindly made the observations at the IAC 80 that allow a relative
calibration of the image taken in the $r^\prime$ filter with the NOT
telescope.

\clearpage

\section*{TABLES}

Table 1. Observations
\begin{center}

\begin{tabular}{llc}
& Imaging \\
\hline
\hline
 Filter & Exp. time (s)  & Seeing  (arcsec) \\
\hline
$u^\prime$ &    2 $\times$ 1800 & 2.7\\
$g^\prime$ &    1 $\times$ 900  & 2.7 \\
$r^\prime$ &    5 $\times$ 1800 & 1.4 \\
\hline
& Spectroscopy \\
\hline
\hline
Position & Exp. time (s)& Slit width\\
\hline
1 & $4 \times 1800$ & 3.0 \\
2 & $1 \times 1800$ & 1.2 \\
3 & $1 \times 1800$ & 1.2 \\
\end{tabular}
\end{center}

\newpage

Table 2. Emission lines

\begin{center}
\begin{tabular}{lrrr}
Line                    &        & EW (\AA) \\
                        &Object 1& Object 2   & Object 3    \\
\hline
\\
OII   ($\lambda 3727$)  &  26    & 28         & 37          \\
NeIII ($\lambda 3869$)  &  17    & 11         &             \\
H$\gamma$               &   6    &  3         &             \\
H$\beta$                &   7    & 14         &  8          \\
OIII ($\lambda 4959$)   &  41    &  2         &  2          \\
OIII ($\lambda 5007$)   & 117    &  8         &  5          \\
NI($\lambda 5198,5200$) &        &  2         &             \\
HeI($\lambda 5876)$     &   1    &  2         &             \\
OI   ($\lambda 6300$)   &   6    &  4         &             \\
NII  ($\lambda 6548$)   &   2    &  8         &  5          \\
H$\alpha$               &  10    & 66         & 52          \\
NII  ($\lambda 6584$)   &   4    & 28         & 17          \\
SII  ($\lambda6 717$)   &   6    & 16         & 10          \\
SII  ($\lambda 6731$)   &   7    & 13         &  3          \\
\end{tabular}
\end{center}

\clearpage
Table 3. Density of objects
\begin{center}
\begin{tabular}{ccr}
Object & $r$ (mag) & N (degree$^{-2}$) \\
\hline
\\
2 & 19.6 & 1500 \\
3 & 20.2 & 2500 \\
4 & 17.3 & 200 \\
\end{tabular}
\end{center}

\newpage

\section*{FIGURE CAPTIONS}

\begin{itemize}

\item []1.A false-colour scale in the $r^\prime$ band of the region around the
system NEQ3 (only a $\sim 40^{\prime\prime}\times 40^{\prime\prime}$ section   has been
plotted). The plot also show
the position of the slits (see main text
for the details). The objects studied in this article are labelled as 1--4.

\item []2. A grey-scale image in the $r^\prime$ band of the three compact
objects, the main galaxy and the filament apparently connecting them. The
contours correspond to 25.6, 26.0 and 26.8 mag arcsec$^{-2}$. North is to the
right and east to the top.

\item []3. Main spectral features (corrected of redshift) of the three
emission line objects analysed in this paper.

\end{itemize}

\newpage
\plotone{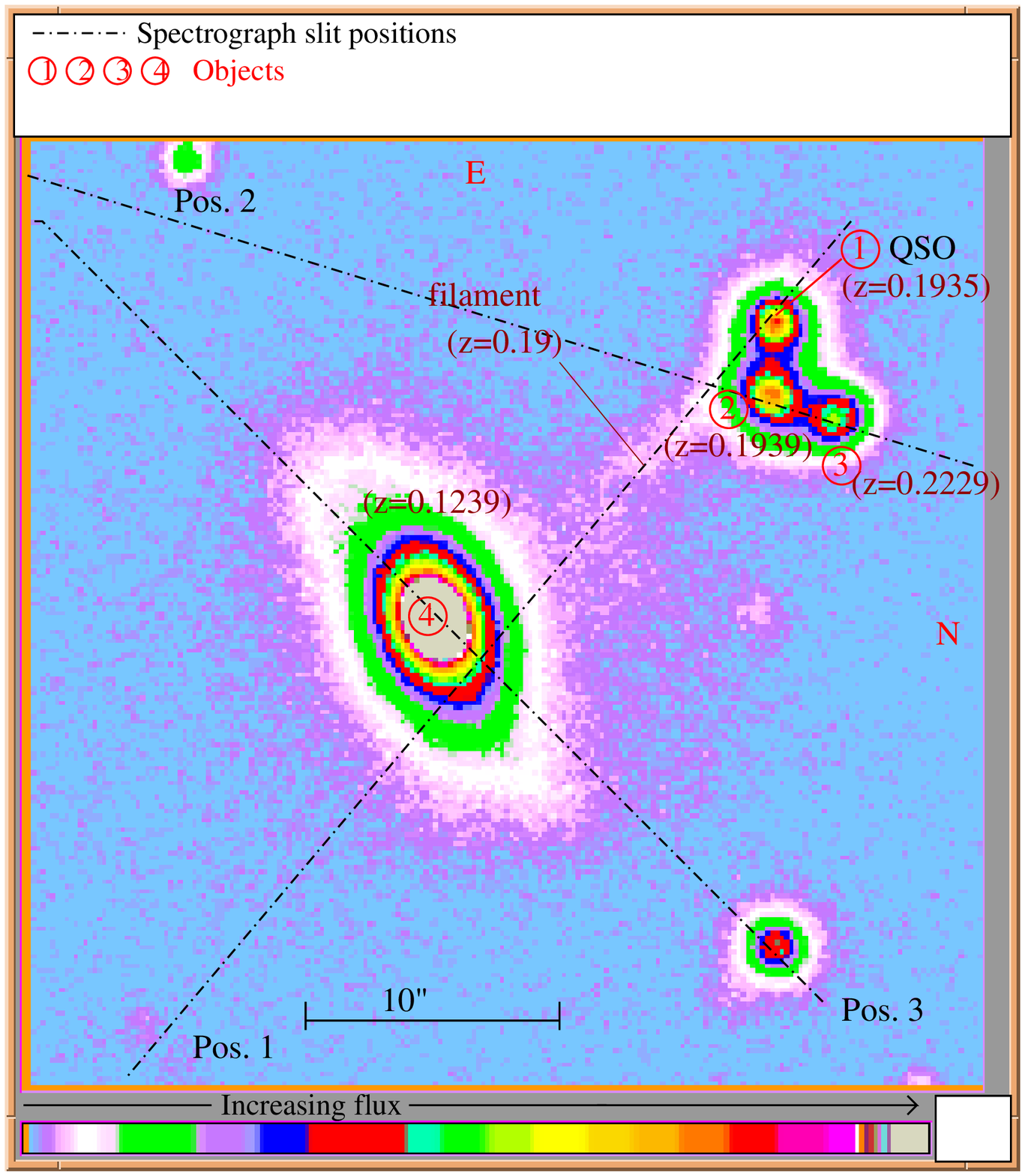}
\newpage
\plotone{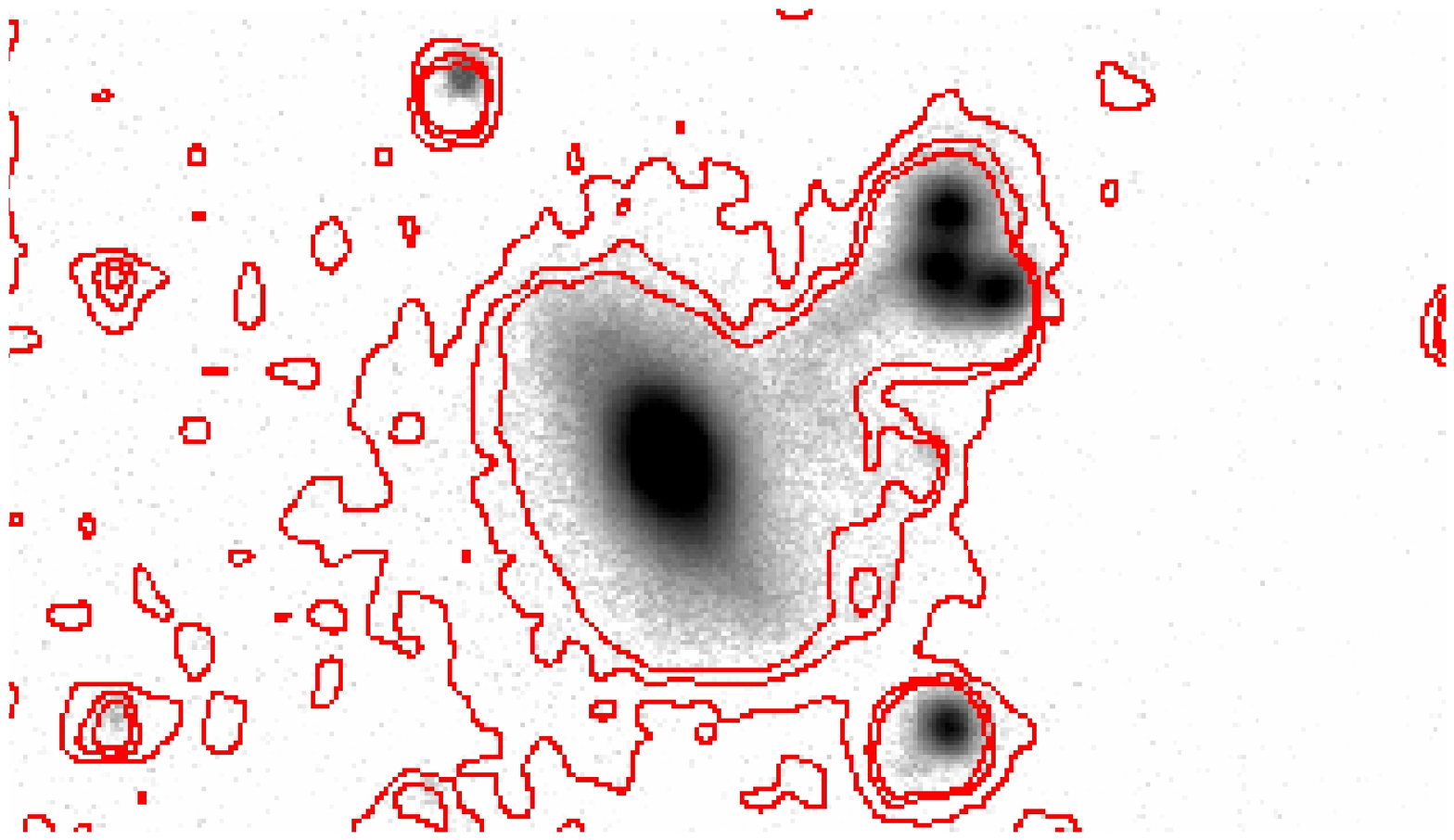}
\newpage
\plotone{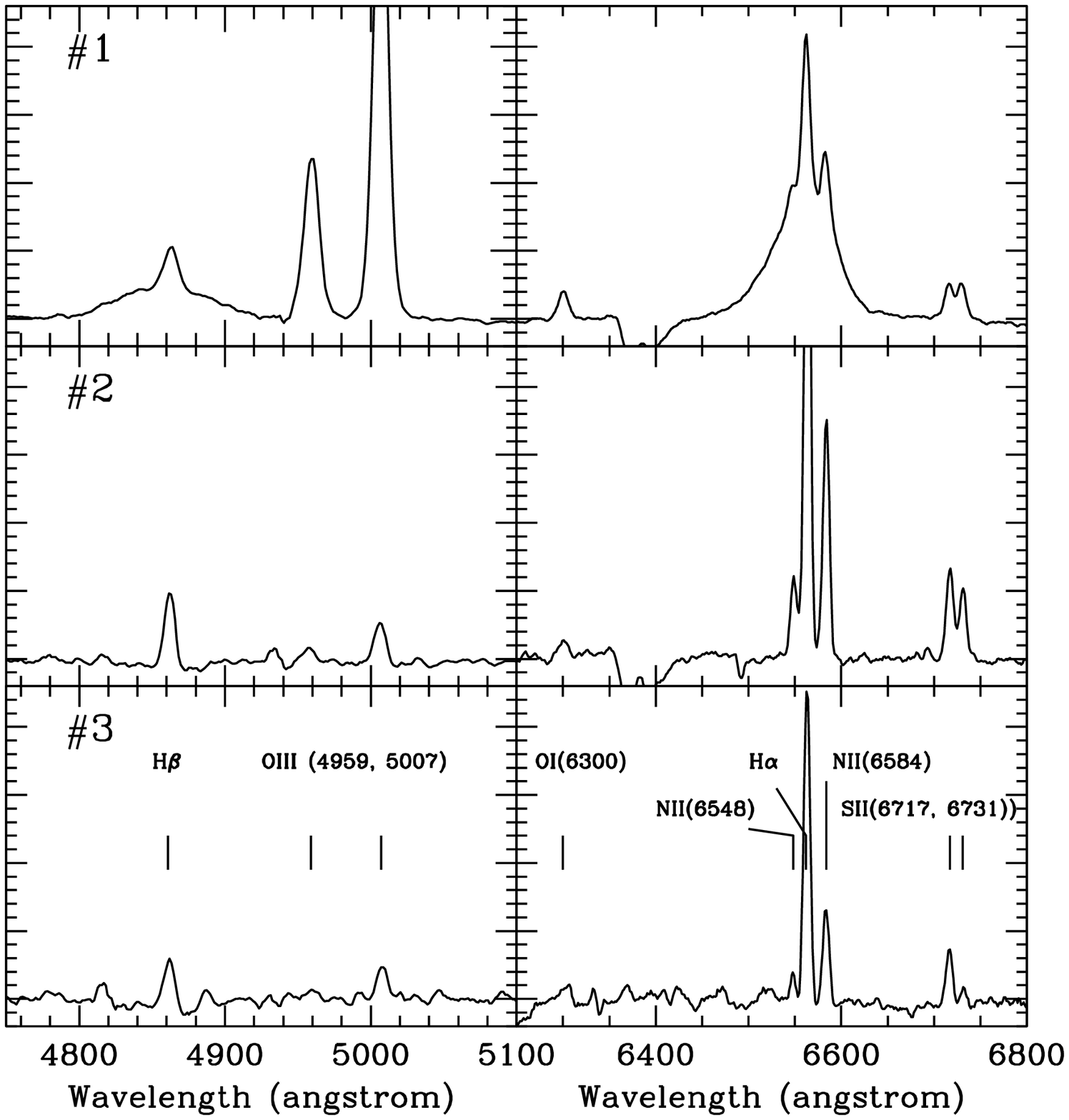}

\end{document}